\documentclass[10pt,english]{article}
\usepackage[T1]{fontenc}
\usepackage[latin1]{inputenc}
\usepackage{geometry}
\usepackage{graphicx}

\makeatletter

\providecommand{\tabularnewline}{\\}

\usepackage{graphicx}

\usepackage{babel}
\makeatother
\begin{document}

\title{\textbf{Optical frequency standard with $\mathrm{Sr^{+}}$: A theoretical
many-body approach}}

\author{Chiranjib Sur , K. V. P. Latha, Rajat K. Chaudhuri, B. P. Das\\
\emph{NAPP Theory Group,} \emph{Indian Institute of Astrophysics,
Bangalore 560 034, India}\\
D. Mukherjee\\
\emph{Indian Association for the Cultivation of Science, Kolkata -
700 032, India}}

\maketitle
\begin{abstract}
Demands from several areas of science and technology have lead to
a worldwide search for accurate optical clocks with an uncertainty
of 1 part in $10^{18}$, which is $10^{3}$ times more accurate than
the present day cesium atomic clocks based on microwave frequency
regime. In this article we discuss the electric quadrupole and the
hyperfine shifts in the $5s\,^{2}S_{1/2}\longrightarrow4d\,^{2}D_{5/2}$
clock transition in $\mathrm{Sr^{+}}$, one of the most promising
candidates for next generation optical clocks. We have applied relativistic
coupled cluster theory for determining the electric quadrupole moment
of the $4d\,^{2}D_{5/2}$ state of $\mathrm{^{88}Sr^{+}}$ and the
magnetic dipole ($A$) and electric quadrupole ($B$) hyperfine constants
for the $5s\,^{2}S_{1/2}$ and $4d\,^{2}D_{5/2}$ states which are
important in the study of frequency standards with $\mathrm{Sr^{+}}$.
The effects of electron correlation which are very crucial for the
accurate determination of these quantities have been discussed. 

~

\textbf{PACS number(s).} : 31.15.Ar, 31.15.Dv, 32.30.Jc, 31.25.Jf,
32.10.Fn 
\end{abstract}

\section{\label{intro}Introduction}

The frequencies at which atoms emit or absorb electro-magnetic radiation
during a transition can be used for defining the basic unit of time
\cite{time-gen1,time-gen2,time-gen3}. The transitions that are extremely
stable, accurately measurable and reproducible can serve as excellent
frequency standards \cite{time-gen1,time-gen2}. The current frequency
standard is based on the ground state hyperfine transition in $\mathrm{^{133}Cs}$
which is in the microwave regime and has an uncertainty of one part
in $10^{15}$ \cite{cs-clock}. However, there is a search for even
more accurate clocks in the optical regime. The uncertainty of these
clocks is expected to be about 1 part in $10^{17}$ or $10^{18}$
\cite{holl}. Some of the prominent candidates that belong to this
category are $\mathrm{^{88}Sr^{+}}$ \cite{bernard,margolis}, $\mathrm{^{199}Hg^{+}}$
\cite{rafac}, $\mathrm{^{171}Yb^{+}}$ \cite{stenger}, $\mathrm{^{43}Ca^{+}}$
\cite{ca+OFS}, $\mathrm{^{138}Ba^{+}}$ etc. Indeed detailed studies
on these ions will have to be carried out in order to determine their
suitability for optical frequency standards. In a recent article \cite{phys-world}
Gill and Margolis have discussed the merits of choosing $\mathrm{^{88}Sr^{+}}$
as a candidate for an optical clock. Till recently, the most accurate
measurement of an optical frequency was for the clock transition in
$\mathrm{^{88}Sr^{+}}$ which has an uncertainty of $3.4$ parts in
$10^{15}$ \cite{margolis-sc04}. However, recently, Oskay et al.
\cite{hg+-prl} have measured the optical frequency of $\mathrm{^{199}Hg^{+}}$
to an accuracy of $1.5$ parts in $10^{15}$ and further improvements
are expected \cite{It-priv}.

\begin{figure}
\begin{centering}\includegraphics{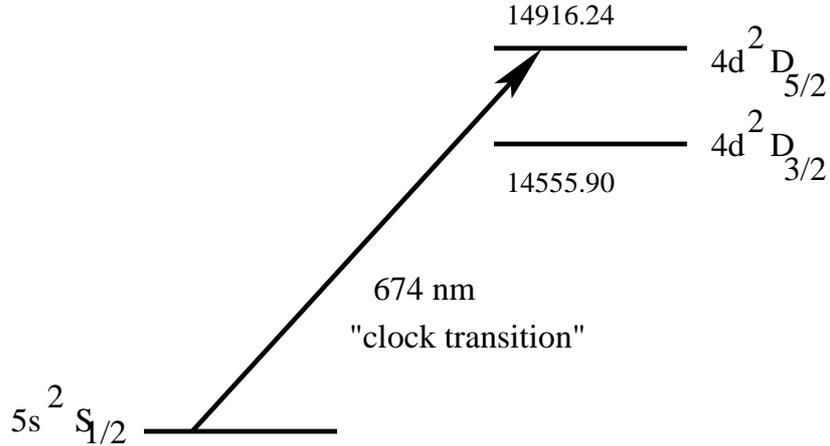}\par\end{centering}

\caption{\label{en-levels}Clock transition in $\mathrm{^{88}Sr^{+}}$. Excitation
energies of the $4d~^{2}D_{3/2}$and $4d~^{2}D_{5/2}$levels are given
in $\mathrm{cm^{-1}}$. }
\end{figure}

In this article we concentrate on strontium ion ($\mathrm{Sr^{+}}$)
which is considered to be one of the leading candidates for an ultra
high precision optical clock \cite{phys-world}. The clock transition
in this case is $5s\,^{2}S_{1/2}\longrightarrow4d\,^{2}D_{5/2}$ and
is observed by using the quantum jump technique in single trapped
strontium ion. When an atom interacts with an external field, the
standard frequency may be shifted from the resonant frequency. The
quality of the frequency standard depends upon the accurate and precise
measurement of this shift. To minimize or maintain any shift of the
clock frequency, the interaction of the atom with it's surroundings
must be controlled. Hence, it is important to have a good knowledge
of these shifts so as to minimize them while setting up the frequency
standard. Some of these shifts are the linear Zeeman shift, quadratic
Zeeman shift, second-order Stark shift, hyperfine shift and electric
quadrupole shift. The largest source of uncertainty in frequency shift
arises from the electric quadrupole shift of the clock transition
because of the interaction of atomic electric quadrupole moment with
the gradient of external electric field. In this article we have applied
relativistic coupled-cluster (RCC) theory, one of the most accurate
atomic many-body theories to calculate the electric quadrupole moment
(EQM) and the hyperfine constants for the energy levels involved in
the clock transition.

\section{\label{RCC}Relativistic coupled-cluster theory }

The relativistic and dynamical electron correlation effects can be
incorporated in many-electron systems through a variety of many-body
methods \cite{grant-scrip,blundell-johnson,OSCC-DM,kaldor}. The relativistic
coupled cluster (RCC) method has emerged as one of the most powerful
and effective tools for high precision description of electron correlations
in many-electron systems \cite{OSCC-DM,kaldor}. The RCC is an all-order
non-perturbative scheme, and therefore, the higher order electron
correlation effects can be incorporated more efficiently than using
the order-by-order diagrammatic many-body perturbation theory (MBPT).
RCC is equivalent to all order relativistic MBPT (RMBPT). The RCC
results can therefore be improved by adding the important omitted
diagrams with the aid of low order RMBPT. We have applied RCC theory
to calculate atomic properties for several systems and more details
can be obtained from Ref. \cite{napp-review}.

\begin{figure}
\begin{centering}\includegraphics{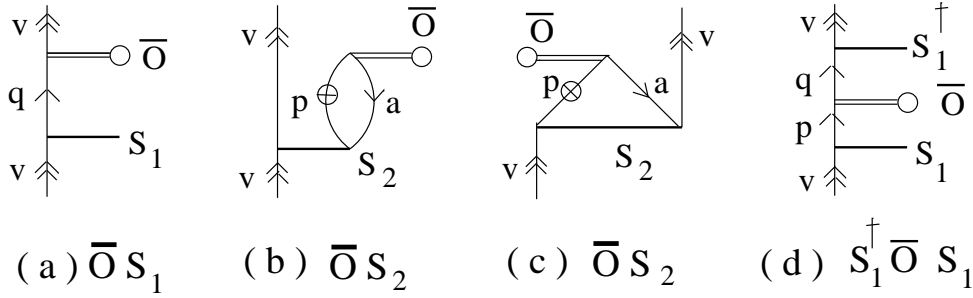}\par\end{centering}

\caption{\label{cc-diag}Some many-body diagrams representing the electric
quadrupole/hyperfine interaction. Holes (occupied orbitals, labeled
by $a$) and particles are denoted by the lines directed downward
and upward respectively. The double line represents the interaction
vertex. The valence (labeled by $v$) and virtual orbitals (labeled
by $p,q,r..$) are depicted by double arrow and single arrow respectively
whereas the orbitals denoted by $\oplus$ can either be valance or
virtual.}
\end{figure}

Here we present a brief outline of RCC theory. We begin with $N$-electron
Dirac-Coulomb Hamiltonian ($H$) which is expressed as

\begin{equation}
H=\sum_{i=1}^{N}\left[c\vec{\alpha_{i}}\cdot\vec{p}_{i}+\beta mc^{2}+V_{N}(r_{i})\right]+\sum_{i<j}^{N}\frac{e^{2}}{r_{ij}}\,,\label{dc}\end{equation}
with the Fermi vacuum described by the four component Dirac-Fock (DF)
state $\left|\Phi\right\rangle $. The normal ordered form of the
above Hamiltonian is given by

\noindent \begin{equation}
H_{N}=H-\langle\Phi|H|\Phi\rangle=\sum_{ij}\langle i|f|j\rangle\left\{ a_{i}^{\dagger}a_{j}\right\} +\frac{1}{4}\sum_{i,j,k,l}\langle ij||kl\rangle\left\{ a_{i}^{\dagger}a_{j}^{\dagger}a_{l}a_{k}\right\} ,\label{eq1}\end{equation}
 where \begin{equation}
\langle ij||kl\rangle=\langle ij|\frac{1}{r_{12}}|kl\rangle-\langle ij|\frac{1}{r_{12}}|lk\rangle.\label{eq2}\end{equation}
 Following Lindgren's formulation of open-shell CC \cite{lindgren-book},
we express the valence universal wave operator $\Omega$ as \begin{equation}
\Omega=\{\exp(\sigma)\},\label{eq7}\end{equation}
 and $\sigma$ being the excitation operator and curly brackets denote
the normal ordering. The wave operator $\Omega$ acting on the DF
reference state gives the exact correlated state. The operator $\sigma$
has two parts, one corresponding to the core and the other to the
valence sector denoted by $T$ and $S$ respectively. In the singles
and double (SD) excitation approximation the excitation operator for
the core sector is given by 

\begin{equation}
T=T_{1}+T_{2}=\sum_{ap}\left\{ a_{p}^{\dagger}a_{a}\right\} t_{a}^{p}+\frac{1}{4}\sum_{abpq}\left\{ a_{p}^{\dagger}a_{q}^{\dagger}a_{b}a_{a}\right\} t_{ab}^{pq}\,,\label{core-T}\end{equation}
$t_{a}^{p}$ and $t_{ab}^{pq}$ being the amplitudes corresponding
to single and double excitations respectively. For a single valence
system like $\mathrm{Sr^{+}}$, the excitation operator for the valence
sector turns out to be $\exp(S)=\left\{ 1+S\right\} $ and

\begin{equation}
S=S_{1}+S_{2}=\sum_{k\neq p}\left\{ a_{p}^{\dagger}a_{k}\right\} s_{k}^{p}+\sum_{bpq}\left\{ a_{p}^{\dagger}a_{q}^{\dagger}a_{b}a_{k}\right\} s_{kb}^{pq}\,,\label{open-S}\end{equation}
where $s_{k}^{p}$ and $s_{kb}^{pq}$ denotes the single and double
excitation amplitudes for the valence sectors respectively. In Eqs.
(\ref{core-T}) and (\ref{open-S}) we denote the core (virtual )
orbitals by $a,b,c...\,(p,q,r...)$ respectively and $k$ corresponds
to the valence orbital. \\
The corresponding correlated closed shell state is then

\begin{equation}
\left|\Psi\right\rangle =\exp(T)\left|\Phi\right\rangle .\label{closed-eqn}\end{equation}
The exact open shell reference state is achieved by using the techniques
of electron attachment. In order to add an electron to the $k$th
virtual orbital of the $N$ electron DF reference state we define

\begin{equation}
\left|\Phi_{k}^{N+1}\right\rangle \equiv a_{k}^{\dagger}\left|\Phi\right\rangle \label{cc-10}\end{equation}
with the particle creation operator $a_{k}^{\dagger}$. Then by using
the excitation operators for both the core and valence electron the
$(N+1)$ electron exact state is defined as \cite{lindgren-book}:

\begin{equation}
\left|\Psi_{k}^{N+1}\right\rangle =\exp(T)\left\{ 1+S\right\} \left|\Phi_{k}^{N+1}\right\rangle .\label{cc-11}\end{equation}
We write the expectation value of any operator $O$ in a normalized
form with respect to the exact state $\left|\Psi^{N+1}\right\rangle $
as

\begin{equation}
\left\langle O\right\rangle =\frac{\left\langle \Psi^{N+1}\right|O\left|\Psi^{N+1}\right\rangle }{\left\langle \Psi^{N+1}\right|\left.\Psi^{N+1}\right\rangle }=\frac{\left\langle \Phi^{N+1}\right|\left\{ 1+S^{\dagger}\right\} \exp(T^{\dagger})O\exp(T)\left\{ 1+S\right\} \left|\Phi^{N+1}\right\rangle }{\left\langle \Phi^{N+1}\right|\left\{ 1+S^{\dagger}\right\} \exp(T^{\dagger})\exp(T)\left\{ 1+S\right\} \left|\Phi^{N+1}\right\rangle }.\label{exp-val}\end{equation}
For computational simplicity we store only the one-body matrix element
of $\overline{O}=\exp(T^{\dagger})O\exp(T)$. $\overline{O}$ may
be expressed in terms of uncontracted single-particle lines \cite{napp-cc}.
The fully contracted part of $\overline{O}$ will not contribute as
it cannot be \emph{linked} with the remaining part of the numerator
of the above equation. First few terms of Eq.(\ref{exp-val}) can
be identified as $\overline{O}$ , $\overline{O}S_{1}$ and $\overline{O}S_{2}$
which we identify as dressed Dirac-Fock (DDF), dressed pair-correlation
(DPC) and dressed core-polarization (DCP) respectively. The other
terms are being identified as , $S_{1}^{\dagger}\overline{O}S_{1}$,
$S_{1}^{\dagger}\overline{O}S_{2}$, $S_{2}^{\dagger}\overline{O}S_{1}$
and $S_{2}^{\dagger}\overline{O}S_{2}$ and are classified as higher
order effects. We use the term `dressed' to describe the modification
of the operator $O$ due to core-core correlation effects. In Fig.
\ref{cc-diag} we replace the operator $O$ by the dressed operator
$\overline{O}$ (the dressed quadrupole/hyperfine interaction operator)
which includes the core excitation effects and the respective figures
are termed as \ref{cc-diag}(a) $\overline{O}S_{1}$, \ref{cc-diag}(b+c)
$\overline{O}S_{2}$ and \ref{cc-diag}(d) $S_{1}^{\dagger}\overline{O}S_{1}$.
Here \ref{cc-diag}(a) and \ref{cc-diag}(b+c) represent the DPC and
DCP effects respectively. \ref{cc-diag}(b) is known as the direct
and \ref{cc-diag}(c) is the exchange DPC diagram. Figure \ref{cc-diag}(d)
refers to one of the higher order pair correlation effect which belongs
to the set, termed as `others'. 

\section{\label{EQS}Electric quadrupole shift}

The largest source of systematic frequency shift for the clock transition
in $\mathrm{Sr^{+}}$ arises from the electric quadrupole shift of
the $4d\,^{2}D_{5/2}$ state caused by its electric quadrupole moment
of that state and the interaction of the external electric field gradient
present at the position of the ion. The electric quadrupole moment
in the state $4d~^{2}D_{5/2}$ was measured experimentally by Barwood
\emph{et al.} at NPL \cite{barwood-srQ}. Since the ground state $5s~^{2}S_{1/2}$
does not possess any electric quadrupole moment, the contribution
to the quadrupole shift for the clock frequency comes only from the
$4d~^{2}D_{5/2}$ state. 

The interaction of the atomic quadrupole moment with the external
electric-field gradient is analogous to the interaction of a nuclear
quadrupole moment with the electric fields generated by the atomic
electrons inside the nucleus. In the presence of the electric field,
this gives rise to an energy shift by coupling with the gradient of
the electric field. Thus the treatment of electric quadrupole moment
is analogous to its nuclear counterpart. The quadrupole moment ${\bf \Theta}$
of an atomic state $|\Psi(\gamma,J,M)\rangle$ is defined as the diagonal
matrix element of the quadrupole operator with the maximum value $M_{J}$,
given by 

\begin{equation}
{\bf \Theta}=\left\langle \Psi(\gamma JJ)\right|\Theta_{zz}\left|\Psi(\gamma JJ)\right\rangle \,.\label{theta}\end{equation}
Here $\gamma$ is an additional quantum number which distinguishes
the initial and final states. The electric quadrupole operator in
terms of the electronic coordinates is given by

\begin{equation}
\Theta_{zz}=-\frac{e}{2}\sum_{j}\left(3z_{j}^{2}-r_{j}^{2}\right),\label{theta-zz}\end{equation}
where the sum is over all the electrons and $z$ is the coordinate
of the $j$th electron. To calculate the quantity we express the quadrupole
operator in its single particle form as 

\begin{equation}
\Theta_{m}^{(2)}=\sum_{m}q_{m}^{(2)}.\label{theta-single}\end{equation}
More details about evaluation of electric quadrupole moment using
RCC theory is described in our recent paper \cite{csur-prl}. The
electric quadrupole shift is evaluated using the relation 

\begin{equation}
\left\langle \Psi(\gamma JFM_{F})\right|\Theta\left|\Psi(\gamma JFM_{F})\right\rangle =\frac{-2A\left[3M_{F}^{2}-F(F+1)\right]\left\langle \Psi(\gamma JF)\right\Vert \Theta^{(2)}\left\Vert \Psi(\gamma JF)\right\rangle }{\left[(2F+3)(2F+2)(2F+1)2F(2F-1)\right]^{1/2}}\times\mathcal{O}(\alpha,\beta)\label{EQshift}\end{equation}
and 

\begin{equation}
\mathcal{O}(\alpha,\beta)=\left[(3\cos^{2}\beta-1)-\epsilon(\cos^{2}\alpha-\sin^{2}\alpha)\right].\label{alpha-beta}\end{equation}
 Here $\gamma$ specifies the electronic configuration of the atoms
and $F$ and $M_{F}$ are the total atomic angular momentum (nuclear
+ electronic) and its projection; $\alpha$ and $\beta$ are the two
of the three Euler angles that take the principal-axis frame of the
electric field gradient to the quantization axis and $\epsilon$ is
an asymmetry parameter of the electric potential function \cite{dube-Q}.

\section{\label{HS}Hyperfine shift}

The frequency standard is based on the $\mathrm{^{88}Sr}$ isotope.
In addition to $\mathrm{^{88}Sr^{+}}$, the odd isotope $\mathrm{^{87}Sr^{+}}$
has also been proposed as a possible candidate for an optical frequency
standard \cite{npl-sr+-hyp}. An experiment has been performed in
NPL to measure the hyperfine structure of the $4d\,\,^{2}D_{5/2}$
state in $\mathrm{^{87}Sr^{+}}$\cite{npl-sr+-hyp}. Theoretical determination
of hyperfine constants is one of most stringent tests of accuracy
of the atomic wave functions near the nucleus. Also accurate predictions
of hyperfine coupling constants require a precise incorporation of
relativistic and correlation effects. 

Unlike the even isotope ($\mathrm{^{88}Sr^{+}}$) of strontium ion,
$\mathrm{^{87}Sr^{+}}$ has a non zero nuclear spin ($I=\frac{9}{2}$)
and the $m_{F}=0$ levels for both the $^{2}S_{1/2}$ and $^{2}D_{5/2}$
states are independent of the first order Zeeman shift. Here, in table
\ref{A and B} we present the results of our calculation of the magnetic
dipole ($A$) hyperfine constant for the $5s\,^{2}S_{1/2}$ and $4d\,\,^{2}D_{5/2}$
states and the electric quadrupole hyperfine constant ($B$) for the
$4d\,\,^{2}D_{5/2}$ state of $\mathrm{^{87}Sr^{+}}$ and compare
with the measured values. More details of our calculation can be found
in Ref \cite{csur-sr-hyp}. 

The hyperfine interaction is given by 

\begin{equation}
H_{hfs}=\sum_{k}M^{(k)}\cdot T^{(k)},\label{HYP-eqn}\end{equation}
where $M^{(k)}$ and $T^{(k)}$ are spherical tensors of rank $k$,
which corresponds to nuclear and electronic parts of the interaction
respectively. The lowest $k=0$ order represents the interaction of
the electron with the spherical part of the nuclear charge distribution. 

In the first order perturbation theory, the energy corresponding to
the hyperfine interaction of the fine structure state $\left|JM_{J}\right\rangle $
are the expectation values of $H_{hfs}$ such that

\begin{equation}
\begin{array}{ccc}
W(J) & = & \left\langle IJFM_{F}\right|{\displaystyle \sum_{k}}M^{(k)}\cdot T^{(k)}\left|IJFM_{F}\right\rangle \\
 & = & {\displaystyle \sum_{k}}(-1)^{I+J+F}\left\{ \begin{array}{ccc}
I & J & F\\
J & I & k\end{array}\right\} \left\langle I\right\Vert M^{(k)}\left\Vert I\right\rangle \left\langle J\right\Vert T^{(k)}\left\Vert J\right\rangle \,.\end{array}\label{hyp-energy}\end{equation}
 Here $\mathbf{I}$ and $\mathbf{J}$ are the total angular angular
momentum for the nucleus and the electron state, respectively, and
$\mathbf{F}=\mathbf{I}+\mathbf{J}$ with the projection $M_{F}$.

\subsection{\label{hyp-A}Magnetic dipole hyperfine constant}

Hyperfine effects arise due to the interaction between the various
moments of the nucleus and the electrons of an atom. Nuclear spin
gives rise to a nuclear magnetic dipole moment which interacts with
the electrons and thus gives rise to magnetic dipole hyperfine interaction
defined by the magnetic dipole hyperfine constant $A$. For an eigen
state $\left|IJ\right\rangle $ of the Dirac-Coulomb Hamiltonian,
$A$ is defined as 

\begin{equation}
A=\mu_{N}\left(\frac{\mu_{I}}{I}\right)\frac{\left\langle J\right\Vert T^{(1)}\left\Vert J\right\rangle }{\sqrt{J(J+1)(2J+1)}},\label{mag-dip}\end{equation}
where $\mu_{I}$ is the nuclear dipole moment defined in units of
Bohr magneton $\mu_{N}$. The magnetic dipole hyperfine operator $T_{q}^{(1)}$
which is a tensor of rank $1$ can be expressed in terms of single
particle rank $1$ tensor operators and is given by the first order
term of Eq. (\ref{hyp-energy})

\begin{equation}
T_{q}^{(1)}=\sum_{q}t_{q}^{(1)}=\sum_{j}-ie\sqrt{\frac{8\pi}{3}}r_{j}^{-2}\overrightarrow{\alpha_{j}}\cdot\mathbf{Y}_{1q}^{(0)}(\widehat{r_{j}})\,.\label{T1}\end{equation}
Here $\overrightarrow{\alpha}$ is the Dirac matrix and $\mathbf{Y}_{kq}^{\lambda}$
is the vector spherical harmonics. The index $j$ refers to the $j$-th
electron of the atom with $r_{j}$ its radial distance and $e$ is
the magnitude of the electronic charge.

\subsection{\label{hyp-B}Electric quadrupole hyperfine constant}

The second order term in the hyperfine interaction is the electric
quadrupole part. The electric quadrupole hyperfine constant is defined
by putting $k=2$ in Eq. (\ref{hyp-energy}). The nuclear quadrupole
moment is defined as

\begin{equation}
T_{q}^{(2)}=\sum_{q}t_{q}^{(2)}=\sum_{j}-er_{j}^{-3}C_{q}^{(2)}(\widehat{r_{j}}),\label{T2}\end{equation}
Here, $C_{q}^{(k)}=\sqrt{\frac{4\pi}{(2k+1)}}Y_{kq}$, with $Y_{kq}$
being the spherical harmonic. Hence the electric quadrupole hyperfine
constant $B$ in terms of the nuclear quadrupole moment $Q_{N}$ is 

\begin{equation}
B=2eQ_{N}\left[\frac{2J(2J-1)}{(2J+1)(2J+2)(2J+3)}\right]^{1/2}\left\langle J\right\Vert T^{(2)}\left\Vert J\right\rangle .\label{Mag-B-eqn}\end{equation}
The corresponding shift in the energy levels are known an hyperfine
shift which is expressed as 

\begin{equation}
W_{hyp}=W_{M1}+W_{E2}=A\frac{K}{2}+\frac{B}{2}\frac{3K(K+1)-4I(I+1)J(J+1)}{2I(2I-1)2J(2J-1)},\label{hyp-shift}\end{equation}
where $K=F(F+1)-I(I+1)-J(J+1)$.

\section{\label{results}Results and discussions}

The occupied and the virtual orbitals used in the calculation are
obtained by solving the Dirac-Fock (DF) equation for $\mathrm{Sr^{++}}$
for a finite Fermi nuclear distribution. These orbitals are linear
combinations of Gaussian type functions on a grid \cite{rajat-gauss}.
The open shell coupled cluster (OSCC) method is used to construct
different single valence states whose reference states correspond
to adding a particle to the closed shell reference state. We use the
singles-doubles and partial triples approximation, abbreviated as
CCSD(T) and excitations from all the core orbitals have been considered.
We have estimated the error incurred in our present work, by taking
the difference between our RCC calculations with singles, doubles
as well as the most important triple excitations (CCSD(T)) and only
single and double excitations (CCSD).

\subsection{\label{quad-result}Electric quadrupole moment }

We present our results of the electric quadrupole moment for the $4d~^{2}D_{5/2}$
state of $\mathrm{^{88}Sr^{+}}$ in table \ref{Quad-result}. The
value of ${\bf \Theta}$ in the $4d~^{2}D_{5/2}$ state measured experimentally
is $(2.6\pm0.3)ea_{0}^{2}$ \cite{barwood-srQ}. Our calculated value
for the $4d~^{2}D_{5/2}$ stretched state is $(2.94\pm0.07)ea_{0}^{2}$,
where $e$ is the electronic charge and $a_{0}$ is the Bohr radius.
We analysed our results and have found that the DDF contribution is
the largest. The leading contribution to electron correlation comes
from the DPC effects and the DCP effects are an order of magnitude
smaller. This can be understood from the DPC diagram (Fig.\ref{cc-diag}(a))
which has a valence electron in the $4d_{5/2}$ state. Hence the dominant
contribution to the electric quadrupole moment arises from the overlap
between virtual $d_{5/2}$ orbitals and the valence, owing to the
fact that $S_{1}$ is an operator of rank 0 and the electric quadrupole
matrix elements for the valence $4d_{5/2}$ and the diffuse virtual
$d_{5/2}$ orbitals are substantial. On the other hand, in the DCP
diagram (Fig.\ref{cc-diag}(b+c)), the matrix element of the same
operator could also involve the less diffuse $s$ or $p$ orbitals.
Hence, for a property like the electric quadrupole moment, whose magnitude
depends on the square of the radial distance from the nucleus, this
trend seems reasonable, whereas for properties like hyperfine interaction
which is sensitive to the near nuclear region, the trend is just the
opposite for the $d$ states \cite{csur-mg+}. As expected, the contribution
of the DHOPC effect i.e., $S_{1}^{\dagger}\bar{O}S_{1}$ (Fig.\ref{cc-diag}(d))
is relatively important as it involves an electric quadrupole matrix
element between the valence $4d_{5/2}$ and a virtual $d_{5/2}$ orbital.

\begin{table}

\caption{\label{Quad-result}Electric quadrupole moment for the $4d~^{2}D_{5/2}$
state of $\mathrm{^{88}Sr^{+}}$ in units of $ea_{0}^{2}$. PW corresponds
to our present CCSD(T) calculation and MCDF for Multi-configuration
Dirac-Fock. }

\begin{centering}\begin{tabular}{llrrl}
\hline 
&
&
PW&
MCDF \cite{itano-pra-Q}&
Experiment \cite{barwood-srQ}\tabularnewline
\hline
$4d_{5/2}$&
&
2.94&
3.02&
2.6$\pm0.3$\tabularnewline
&
&
&
&
\tabularnewline
\hline
\hline 
&
&
&
&
\tabularnewline
\end{tabular}\par\end{centering}
\end{table}

\subsection{\label{hyp-result}Hyperfine constants}

The magnetic dipole $(A$) and electric quadrupole $(B$) hyperfine
constants for the of $\mathrm{^{87}Sr^{+}}$ are given in table \ref{A and B}
along with the calculated and experimental results. The $g_{i}=\mu_{N}\left(\frac{\mu_{I}}{I}\right)$
value used for the calculation is from Ref. \cite{web-elm}. PW corresponds
to the `present work' using CCSD(T). CC stands for coupled-cluster
calculation by Martensson-Pendrill \cite{pendrill} and Nayak and
Chaudhuri \cite{malaya}, DF-AO for Dirac-Fock with all order core
polarization effect by Yu \emph{et al.} and the column `others' refers
to the calculation by Yuan \emph{et al}. using relativistic linked
cluster many-body perturbation theory (RLCMBPT) \cite{yuan} and by
one of the authors using relativistic effective valence shell Hamiltonian
method \cite{rajat-freed}. Information on $A$ for the $4d\,\,^{2}D_{5/2}$
state is very important in connection with optical frequency standards
\cite{margolis-sc04,npl-sr+-hyp}. The measured value of this quantity
is $2.1743\pm0.0014$ MHz \cite{npl-sr+-hyp}, whereas the previously
calculated values vary from $1.07$ MHz \cite{pendrill} and $2.507$
MHz \cite{sr-china}. Our calculated value of $A$ is $2.16\pm0.02$
MHz; this is the most accurate theoretical determination of $A$ for
the $4d\,\,^{2}D_{5/2}$ state to date. For the $5s\,\,^{2}S_{1/2}$state
our calculated value of $A$ is $997.26\pm0.03$ MHz. To analyze the
result we focus on the various many-body effects contributing to the
calculation of $A$. The most important many-body diagrams are presented
in Fig. \ref{cc-diag}. We have noticed that for $5s\,\,^{2}S_{1/2}$
state the dominant contribution is at the DDF level $\sim72\%$. However,
for the $4d\,\,^{2}D_{5/2}$ state the DCP effect is larger and its
sign is opposite that of the DDF contribution. In their calculations
Martensson \cite{pendrill} and Yu \emph{et al.} \cite{sr-china}
have found similar trends. We have seen that the higher order correlation
effects contribute significantly in determining $A$ for this state
- collectively they are $~60\%$ of the total value but opposite in
sign. In the earlier calculations the determination of the higher
order effects was not as accurate as ours. 

The calculated value for the electric quadrupole hyperfine constant
($B$) for the $4d\,\,^{2}D_{5/2}$ state is $47.8\pm0.2$ MHz which
deviates $\sim2.7\%$ from the central experimental value. The earlier
determination of $B$ was off by $\sim11\%$ from the experiment.
Since the other state involved in the clock transition is spherically
symmetric the electric quadrupole hyperfine constant is zero and there
will be no hyperfine shift due to $B$ for the $5s\,\,^{2}S_{1/2}$state.

\begin{table}

\caption{\label{A and B}Magnetic dipole ($A$) hyperfine constant for the
$5s\,^{2}S_{1/2}$ and $4d\,\,^{2}D_{5/2}$ states and the electric
quadrupole hyperfine constant ($B$) for $4d\,\,^{2}D_{5/2}$ state
of $\mathrm{^{87}Sr^{+}}$ in MHz. }

\begin{centering}\begin{tabular}{llrrrll}
\hline 
&
&
PW&
CC &
DF-AO \cite{sr-china}&
Others&
Experiment\tabularnewline
\hline
\hline 
$5s_{1/2}$&
$A$&
-997.26&
-1000 \cite{pendrill}&
-1003.18&
-987 \cite{yuan}&
-1000.5$\pm$1.0 \cite{sr-exp-1992}\tabularnewline
&
&
&
-999.89 \cite{malaya}&
&
-1005.74 \cite{rajat-freed}&
-990 \cite{beigang-83}\tabularnewline
&
&
&
&
&
&
-993.5 \cite{lu-88}\tabularnewline
&
&
&
&
&
&
-1000.473 673 \cite{sunaoshi}\tabularnewline
&
&
&
&
&
&
\tabularnewline
$4d_{5/2}$&
$A$&
2.16&
1.07 \cite{pendrill}&
2.51&
&
2.1743$\pm$0.0014 \cite{npl-sr+-hyp}\tabularnewline
&
&
&
1.87 \cite{malaya}&
&
&
\tabularnewline
&
$B$&
47.8&
54.4 \cite{pendrill}&
&
&
49.11$\pm0.06$ \cite{sr-exp-1992} \tabularnewline
&
&
&
51.12 \cite{malaya}&
&
&
\tabularnewline
\hline
\hline 
&
&
&
&
&
&
\tabularnewline
\end{tabular}\par\end{centering}
\end{table}

From the figure presented in Fig. \ref{cc-diag} it is clear that
the DPC effect involves the hyperfine interaction of a valence electron
and the residual Coulomb interaction, i.e for $5s\,\,^{2}S_{1/2}$
state the hyperfine matrix element becomes $\left\langle [4p^{6}]5s_{1/2}\left|h_{hfs}\right|[4p^{6}]q\right\rangle $,
where $q$ can be any virtual orbital and $h_{hfs}$ is the single-particle
hyperfine operator. Since only $s_{1/2}$ and $p_{1/2}$ electrons
have a sizable density in the nuclear region, the DPC effect is dominant
for $5s_{1/2}$ state but not for the $4d_{5/2}$ state. On the other
hand, the DCP effects represent the hyperfine interaction of a polarized
core electron with any virtual electron (see Fig. \ref{cc-diag}b,c).
For $4d_{5/2}$ state it is clear that the hyperfine matrix element
for the DPC effect is much smaller than the DCP effect, which plays
the most important role in determining the value of $A$. The core
polarization contribution is so large for this state that it even
dominates over the DDF contribution.

\section{\label{conclusion}Conclusion}

In conclusion, we have performed an \emph{ab initio} calculation of
the electric quadrupole moment for the $4d~^{2}D_{5/2}$ state of
$\mathrm{^{88}Sr^{+}}$ to an accuracy of less than $2.5\%$ using
the RCC theory. This is the first application of RCC theory to determine
the electric quadrupole moment (EQM) of any atom and is currently
the most accurate determination of EQM for the $4d~^{2}D_{5/2}$ state
in $\mathrm{Sr^{+}}$. We have also calculated the magnetic dipole
($A$) hyperfine constant for the $5s~^{2}S_{1/2}$ and $4d~^{2}D_{5/2}$
states and electric quadrupole hyperfine constant ($B$) for the $4d~^{2}D_{5/2}$
state of $\mathrm{^{87}Sr^{+}}$. Evaluation of correlation effects
to all orders as well as the inclusion of the dominant triple excitations
in our calculation was crucial in achieving this accuracy. The magnitude
of electric quadrupole moment depends on the square of the radial
distance from the nucleus, whereas properties like hyperfine interaction
are sensitive to the near nuclear region. The accurate determination
of quantities like electric quadrupole moment and hyperfine constants
establish the fact that RCC is very powerful and efficient method
for determining atomic properties near the nuclear region as well
as at large distances from the nucleus. Our result will lead to a
better quantitative understanding of the electric quadrupole shift
of the resonance frequency of the clock transition in $\mathrm{Sr^{+}}$. 

\begin{verse}
\textbf{Acknowledgments} : Financial support from the BRNS, DAE for
project no. 2002/37/12/BRNS is gratefully acknowledged. The computations
are carried out in our group's Xeon and the Opteron computing cluster
at IIA. 
\end{verse}

\end{document}